\documentclass[aps,prd,onecolumn,groupedaddress,nofootinbib,showpacs,showkeys,amssymb]{revtex4}
\usepackage[colorlinks,citecolor=blue,urlcolor=blue,linkcolor=blue]{hyperref}
\newlength{\extraspace}
\newlength{\extraspaces}
\begingroup
\endgroup

\newcommand{\be}{\begin{equation}
\addtolength{\abovedisplayskip}{\extraspaces}
\addtolength{\belowdisplayskip}{\extraspaces}
\addtolength{\abovedisplayshortskip}{\extraspace}
\addtolength{\belowdisplayshortskip}{\extraspace}}
\newcommand{\ee}{\end{equation}}

\newcommand{\ba}{\begin{eqnarray}
\addtolength{\abovedisplayskip}{\extraspaces}
\addtolength{\belowdisplayskip}{\extraspaces}
\addtolength{\abovedisplayshortskip}{\extraspace}
\addtolength{\belowdisplayshortskip}{\extraspace}}
\newcommand{\ea}{\end{eqnarray}}

\newcommand{\nonu}{\nonumber \\[.5mm]}
\newcommand{\A}{&\!\!\!}

\newcommand{\newsection}[1]{
\vspace{7mm} \pagebreak[3] \addtocounter{section}{1}
\setcounter{subsection}{0} 
\begin{center}
{\large {\bf \thesection. #1}}
\end{center}
\nopagebreak
\medskip
\nopagebreak \hspace{3mm}}

\begin{document}

\title{ Charged Anti-de Sitter   BTZ black holes in Maxwell-$f(T)$ gravity}

\author{  G.  G. L. Nashed$^{1,2}$ and S. Capozziello$^{3,4,5}$}

\affiliation{$ ^1$Centre for Theoretical Physics, The British University in Egypt, P.O. Box 43,\\ El Sherouk City, Cairo 11837, Egypt\\
$^2$Department of Mathematics, Faculty of Science, Ain Shams University, Cairo 11566, Egypt\\
$^3$Dipartimento di Fisica ``E. Pancini``, Universit\'a di Napoli ``Federico II'',
Complesso Universitario di Monte Sant' Angelo, Edificio G, Via Cinthia, I-80126, Napoli, Italy\\
$^4$ Istituto Nazionale di Fisica Nucleare (INFN),  Sezione di Napoli,
Complesso Universitario di Monte Sant'Angelo, Edificio G, Via Cinthia, I-80126, Napoli, Italy\\
$^5$Gran Sasso Science Institute, Viale F. Crispi, 7, I-67100, L'Aquila, Italy.\\
}
\begin{abstract}
Inspired by  the  BTZ formalism, we discuss the  Maxwell-$f(T)$ gravity in  (2+1)-dimensions. The main task is to derive exact solutions for a special form of $f(T)=T+\epsilon T^2$, with $T$ being the torsion scalar of Weitzenb$\ddot{\mbox{o}}$ck  geometry. To this end, a triad field is applied to the equations of motion of charged $f(T)$ and sets of circularly symmetric non-charged and charged solutions have been derived.  We show that, in the charged case,  the {\it monopole}-like and the ${\it ln}$ terms are linked by a correlative constant despite of known results  in teleparallel  geometry and its extensions [39].  Furthermore, it is possible to  show that the event horizon is not identical with the Cauchy horizon due to such a constant. The singularities and the horizons of these  black holes are examined: they are new and have no analogue in  literature due to the fact that their curvature singularities are soft.  We calculate  the energy content of these solutions by using the general vector form of the energy-momentum within the framework of $f(T) $ gravity. Finally,  some thermodynamical quantities, like  entropy and Hawking temperature, are derived.
\end{abstract}
\date{\today}

\maketitle

\begin{center}
\newsection{\bf Introduction}
\end{center}
The explanation of the gravitational phenomena at large scales  is difficult  and it is considered as one of the main  issues in physics.
 For example, the  accelerated phase of the universe that is  observationally probed cannot be investigated in the framework of General Relativity (GR) except by introducing  a cosmic fluid possessing  exotic characteristics, like   dark energy,  or inserting the cosmological constant that gives rise to  other conceptual problems  \cite{Ra}-\cite{Pp}. In the same manner, the rotation curves of galaxies  appear to depart  from the standard  gravitational  behavior asking for a large amount of dark matter   \cite{BT}.

 Despite of these issues, GR has attained  excellent achievements in describing the gravitational field in the last 100 years. The accuracy of GR  is checked when theoretical predications and observations  are challenged.  However, the searching for a self-consistent gravitational theory at all scales  is still an open question. A reliable gravitational theory should be able to trace gravitational fields in all domains, in addition,  due to the fact that  GR is consistent  at Large scales with  observations,  any new reliable gravitational theory must tend to GR in a  suitable limit \cite{Be}.

One of a possible way out to the above problems is to extend Einstein's GR   on a  geometric background. Many theories of the gravitational field possessing  variety of geometric formulations have been recently  built up: for example,  $f(R)$ gravity that relies on arbitrary    functions  of the   Ricci scalar  \cite{CF,CDL,NO,NOO}.  Assuming  $f(R) = R$,  the Einstein-Hilbert Lagrangian,  and thus GR, is recovered. Another reliable gravitation theory  is the one that comes from  the generalization of the Weitzenb\"ock geometry, i.e., teleparallel equivalent of Einstein GR  (TEGR) \cite{Wc,N08,APB,Mj}.  The TEGR  is built on the   Riemann-Cartan geometry  where a non-symmetric Weitzenb\"ock connection  is defined: it gives rise to a vanishing curvature and a non-vanishing  torsion.  In TEGR, we can deal with torsion tensor as the key ingredient instead of curvature, whilst the tetrad (4-dimension) field is considered as the dynamical quantity  alternative to  the metric one.   It is  interesting to mention   that Einstein himself,  in his  attempt to  unify gravitational and electromagnetic fields, used TEGR   \cite{Ea,N06,N008,Ea28,Ea29}. Despite of the fact that GR and TEGR  are gravitational theories  having  different geometric structures,  they give rise to  identical field equations and  are invariant under local Lorentz transformations.   Therefore, we can consider that any solution  satisfying the field equations of  GR is also a solution satisfying  the field equations of TEGR. On the other hand, the straightforward extension of TEGR,  the  $f(T)$ gravity,  consists in   Lagrangians depending on  functions of the scalar torsion $T$ \cite{BF9, noi}. Such gravitational theories are  attractive from many aspects. Firstly,  they cannot be  directly  matched  to GR \cite{FF,FF09}. This means that $f(T)$ is not a simple analogue of $f(R)$ in the case of torsion \cite{noi}.  Secondly,  they can be viewed as talent theories to solve several   problems of GR  as explained, for example, in  \cite{CRC}-\cite{Ngrg15}. Due to this versatility,  many studies  on $f(T)$ gravity  have been done, ranging from  exact cosmological solutions to stellar models \cite{ZA}--\cite{CADT}. There is a  price to pay in the  approach  of $f(T)$ which  is the fact that this theory is not invariant under local Lorentz transformations, and therefore different tetrads could arise different field equations \cite{LSB,LSB1}. It is an important issue to note that $f(T)$ theory is a frame-dependent because any solution of its equation of motion depends on the tetrad \cite{LSB,LSB1,BC17}. However,  we can forget this problem and discuss solutions in the special tetrad, this is like what happens in the electromagnetism when one
study  the special class of inertial frames  \cite{KS}.

At a fundamental level, the (3+1) formalism, working in GR, has to be developed also in TEGR and its extension $f(T)$, in view of achieving a consistent quantization approach. In fact,
it is believed  that the (3+1)-dimensional formulation of GR is  one of the best formulation of gravitational field, however, its quantization shows many  problems. Due to these shortcomings, the (2+1)-dimensional formulation of   gravity has accomplished much interest, because  classically it is easy to deal with it and  one can explain  in  more efficiency  a quantization procedure.
Ba$\tilde{\mbox{n}}$ados, Teitelboim and Zanelli (BTZ), in (1992), showed that there is a  solution corresponding to 3-dimensional GR which has a negative value of the cosmological constant \cite{BTZ}.  BTZ  solution  shows several interesting characteristics ranging  from  classical to quantum levels; for example,  some interesting contributes to the Kerr black-hole in (3+1)-dimensions of GR have been developed starting from BTZ result \cite{Cs,Cs1}.

 Actually,   among the motivations that make (2+1)-dimension gravity a remarkable toy model, there  is the existence of the BTZ solution. It has been proved that the BTZ black hole arises from collapsing matter \cite{RM}. This kind of black hole    requires a constant curvature in local spacetime \cite{HW}. In fact, it has been shown that for a certain subset of Anti-de Sitter (AdS) spacetimes \cite{BHTZ}, there is a solution   which one can consider as a black hole. Also, a charged BTZ  solution,  arising  form  AdS-Maxwell gravity in (2+1)-dimensions, has been derived \cite{Cs,MTZ,Cg}; 3-dimensional dilatonic solutions, using a nonlinear electrodynamics,  has been studied in \cite{HPPS}. It is interesting to note that a 3-dimensional  charged black hole has been discussed using the quadratic  form of $f(T)$ \cite{GSV}.

Nevertheless  the  studies in 3-dimensions, the final formulation  of a self-consistent  quantum  gravity  theory is still an open question. Thus, it is interesting to go deep in  3-dimension outlines to check the features, as a preliminary  step to  investigate the (3+1)-dimension gravity.

The main goal of the present paper is  deriving   rotating non-charged and charged black hole solutions in the 3-dimensional   Maxwell-$f(T)$  gravity. This leads to solutions  which  asymptotically behave as AdS black holes for the special quadratic form of $f(T)$. Among the advantages of these solutions, there is the fact that the  electric potentials  has a monopole term, in addition to the logarithmic term,  which  are correlated by a constant. The second term exists  despite of the fact that  these solutions have a singularity  when the radial coordinate is vanishing, i.e., $r=0$:  however, this singularity is much softer than any AdS charged or non-charged solutions  derived in the framework of GR or TEGR \cite{Nijgmmp}. Finally, besides the fact that these solutions  behave asymptotically as AdS,  they have distinct spatial and temporal components, i.e.   $g_{tt}$ and $g^{rr}$ components  are different and  have different  event and Killing horizons.

The outline of the paper is the   following. In \S 2, the Maxwell-$f(T)$  gravity is sketched. In \S 3, a triad field having 3 unknown functions is provided and applied  to the non-charged and charged field equations of $f(T)$ gravity. New  exact non-charged and charged solutions are derived also in \S 3. In \S 4, the physics of the   solutions is  discussed by   discussing the singularities of the scalars constructed from the  Levi-Civita connection and from  the Weitzenb$\ddot{\mbox{o}}$ck. Furthermore, in \S  4, we derive the total energy  related to each solution pointing out  the physical meaning of the integration constants.  In \S 5, some  thermodynamical quantities are discussed. We  show that the first law of thermodynamics is not satisfied for the charged black hole. Final section is devoted to  some concluding remarks.

\section{ The Maxwell-$f(T)$ gravity}\label{S2}
\subsection{The Weitzenb\"{o}ck geometry }\label{S1.1}
The Weitzenb\"{o}ck geometry is assigned  by the couple $\{{\cal M},~ h_{i}\}$, where ${\cal  M}$ is an  N-dimensional manifold and $ h_{i}$ ($ i=1,2,\cdots, N$) are $N$ vectors   globally defined  on the  manifold ${\cal M}$. The vectors  $ h_{i}$ are called the parallelization  fields. In  N-dimensions,   the covariant derivative of the  covariant tetrad field  is vanishing, that is
\begin{equation}\label{q1}
 {h^i}_{\mu; \nu}\stackrel {\rm def.}{=}\partial_{\nu}
{h^i}_\mu-{\Gamma^\lambda }_{\mu \nu} {h^i}_\lambda=0,
\end{equation}
where  $";"$ represents the covariant derivative and the ordinary derivative $","$ is defined as  $\partial_{\nu}\stackrel {\rm def.}{=}\frac{\partial}{\partial x^{\nu}}$. The connection  ${\Gamma^\lambda }_{\mu \nu}$ is the Weitzenb\"{o}ck non-symmetric connection    \cite{Wr}  has the form
\begin{equation}\label{q2}
{\Gamma^\lambda}_{\mu \nu} \stackrel {\rm def.}{=}{h_i}^\lambda~ \partial_\nu h^{i}{_{\mu}}.
\end{equation}
The  tensor $g_{\mu \nu}$ is defined as
\begin{equation}\label{q3}
 g_{\mu \nu} \stackrel {\rm def.}{=}  \eta_{i j} {h^i}_\mu {h^j}_\nu,
\end{equation}
which is the metric tensor with $\eta_{i j}=(+,-,-,- \cdots)$ being  the Minkowskian spacetime. The condition of  metricity is satisfied  as a consequence of Eq. (\ref{q1}). Eq. (\ref{q2}) has  an interesting property that it gives a vanishing curvature tensor  and a non-vanishing torsion tensor.  We note that the tetrad field ${h_i}^\mu$ fixes a unique metric $g^{\mu \nu}$  while the inverse statement is not correct. The torsion and the contortion tensors are defined as
\begin{eqnarray}
\nonumber {{\rm T}^\alpha}_{\mu \nu}  & \stackrel {\rm def.}{=} &
{\Gamma^\alpha}_{\nu \mu}-{\Gamma^\alpha}_{\mu \nu} ={h_i}^\alpha
\left(\partial_\mu{h^i}_\nu-\partial_\nu{h^i}_\mu\right),\\
{{\rm K}^{\mu \nu}}_\alpha  & \stackrel {\rm def.}{=}&
-\frac{1}{2}\left({{\rm T}^{\mu \nu}}_\alpha-{T^{\nu
\mu}}_\alpha-{{\rm T}_\alpha}^{\mu \nu}\right). \label{q4}
\end{eqnarray}
The teleparallel torsion scalar of TEGR theory is defined as
\begin{equation}\label{Tor_sc}
{\rm T}\stackrel {\rm def.}{=}{{\rm T}^\alpha}_{\mu \nu} {{\rm S}_\alpha}^{\mu \nu},
\end{equation}
where the tensor ${{\rm S}_\alpha}^{\mu \nu}$  is  anti-symmetric  in the last two pairs and  has the form
\begin{equation}\label{q5}
{{\rm S}_\alpha}^{\mu \nu}\stackrel {\rm def.}{=}\frac{1}{2}\left({{\rm K}^{\mu\nu}}_\alpha+\delta^\mu_\alpha{{\rm T}^{\beta
\nu}}_\beta-\delta^\nu_\alpha{{\rm T}^{\beta \mu}}_\beta\right).
\end{equation}
Using Eq. (\ref{q4}). Eq. (\ref{q2}) can be rewritten as
\begin{equation}\label{contortion}
    {\Gamma^\mu}_{\nu \rho }=\left \{_{\nu  \rho}^\mu\right\}+{{\rm K}^{\mu}}_{\nu \rho},
\end{equation}
 where $\left \{_{\nu  \rho}^\mu\right\}$  is the Levi-Civita connection of GR theory, that  depends on   $g_{\mu \nu}$ as well as its first derivatives, while  ${{\rm K}^{\mu}}_{\nu \rho}$ is the contortion tensor  that depends on the tetrad fields ${h_i}^\mu$ as well as its first derivatives.
\subsection{The Maxwell-$f(T)$ gravitational theory}\label{S1.2}
Using the same approach as for  $f(R)$ gravity,  we define an arbitrary analytic function  of the
scalar torsion $T$, i.e.,  $f(T)$ gravitational theory in the 3-dimension  action as:
\begin{equation}\label{q7}
{\rm {\cal L}}=\frac{1}{2\kappa_3}\int |h|(f(T)-2\Lambda)~d^{3}x+\int |h|{\cal L}_{ em}~d^{3}x,
\end{equation}
where  $\kappa_3$  is a three-dimensional  constant and $\Lambda$ being the cosmological constant. In Eq. (\ref{q7}) $ |h|=\sqrt{-g}=\det\left({h^a}_\mu\right)$ and ${\cal L}_{
em}=-\frac{1}{2}{ F}\wedge ^{\star}{F}$ is the Maxwell Lagrangian with $F = dA$ and $A=A_{\mu}dx^\mu$ being the electromagnetic
 gauge potential \cite{CGSV13}. Making the variation of Eq. (\ref{q7}) with respect to the triad ${h^i}_\mu$ and the gauge potential gives \cite{BF9,CGSV13,ACN}
\begin{eqnarray}\label{q8}
& & {\rm {S_\mu}^{\rho \nu} \partial_{\rho} T f_{TT}}+\left[h^{-1}{h^i}_\mu\partial_\rho\left(h{h_i}^\alpha
{S_\alpha}^{\rho \nu}\right)-{T^\alpha}_{\lambda \mu}{S_\alpha}^{\nu \lambda}\right]f_T
-\frac{f-2\Lambda}{4}\delta^\nu_\mu +\kappa_3\Theta_\mu{}^\nu=I_\mu{}^\nu\equiv0,\nonumber\\
&&{\rm \partial_\nu \left( \sqrt{-g} F^{\mu \nu} \right)}=0.
\end{eqnarray}
with $f \stackrel {\rm def.}{=}f(T)$, \ \   $f_{T}\stackrel {\rm def.}{=}\frac{\partial f(T)}{\partial T}$, \ \  $f_{TT}\stackrel {\rm def.}{=}\frac{\partial^2 f(T)}{\partial T^2}$.  $\Theta_\mu{}^\nu$ is the
energy-momentum tensor defined as
\be\Theta_\mu{}^\nu=F_{\mu \alpha}F^{\nu \alpha}-\frac{1}{4} \delta_\mu{}^\nu F_{\alpha \beta}F^{\alpha \beta}.\ee   Now we are going
to  study two separate cases individually,  the case of vacuum as well as the case of non-vacuum case respectively.

Eq. (\ref{q8}) can be has the form
\be \partial_\alpha \Biggl[h{S}^{b \beta \alpha} f(T)_T\Biggr]=\kappa_3 h
{h^b}_\mu \Biggl[\tau^{\beta \mu}+\Theta^{\beta \mu}\Biggr],\ee
with $\tau^{\nu \mu}$  being defined as \be \tau^{\nu
\mu}\stackrel {\rm def.}{=}\frac{1}{\kappa_3{}^2}\Biggl[4f(T)_T {S^\alpha}^{\nu
\lambda}{T_{\alpha \lambda}}^{\mu}-g^{\nu \mu} f(T)\Biggr].\ee  Because of  the skewness of  ${S}^{a \nu \lambda}$ we have \be\label{q9}
\partial_\alpha \partial_\beta\left[h{S}^{a \alpha \beta} f_T\right]=0, \quad
\mbox{ which leads to} \qquad
\quad \partial_\beta\left[h\left(\tau^{b \beta}+\Theta^{b
\beta}\right)\right]=0. \ee From Eq. (\ref{q9}) we get \be \label{q10}
\frac{d}{dt}\int_\Sigma d^2x \ h \ {h^a}_\mu \left(\tau^{0 \mu}+\Theta^{0
\mu}\right)+\oint_C \left[h \ {h^a}_\mu \ \left(\tau^{j
\mu}+\Theta^{j
\mu}\right)\right]{\hat n}\cdot dl=0,\ee where $C$ is a contour enclosing the surface $\Sigma$,  ${\hat n}$  is a unit normal  vector to
the closed contour $C$, and $dl$ is an infinitesimal length. Eq. (\ref{q10})  gives the conservation  of
 the energy-momentum and of the quantity
$\tau^{\lambda \mu}$.   Hence,
the total energy-momentum  of (2+1)-dimensional $f(T)$  theory contained
in two-dimensional surface $\Sigma$ is defined as \be \label{q11} P^b:=\int_\Sigma d^2x
\ h \ {h^b}_\mu \left(\tau^{0 \alpha}+\Theta^{0
\alpha}\right)=\frac{1}{\kappa_3}\int_\Sigma d^2x  \partial_\alpha\left[h{S}^{b 0
\alpha} f(T)_T\right].\ee Eq. (\ref{q11}) is   the generalization of the energy-momentum tensor for the $f(T)$ theory.The above equation  can be used to carry out the calculation of energy and momentum and, as soon as  $f(T)=T$,
it returns to the well know form  of  (2+1)-dimensional TEGR \cite{MDTC}.

It is important to mention here  that Eq. (\ref{q10}) is valid only for  solutions which behave asymptotically as a flat spacetime; however, for  solutions which behave  asymptotically as AdS/dS, Eq. (\ref{q10}) is not valid because the second term will not vanish asymptotically. Therefore, we must add a quantity  which assures the vanishing of the second term asymptotically for any  solution  which behaves as Ads/dS. This expression has the form $T^{\mu \nu}$ and, in that case, Eq. (\ref{q10}) takes the form
 \be
\frac{d}{dt}\int_\Sigma d^2x \ h \ {h^a}_\mu \left(\tau^{0 \mu}+\Theta^{0
\mu}\right)+\oint_C \left[h \ {h^a}_\mu \ \left(\tau^{j \mu}+\Theta^{j \mu}+T^{j \mu}\right)\right]{\hat n}\cdot dl=0,\ee
with $T^{j  \mu}$ being the energy-momentum  of pure AdS/dS spacetime.
\newsection{Three-dimensional black holes in Maxwell-$f(T)$ gravity}
Using the coordinate $\{t, r, \phi\}$, we write  the triad  that possesses three unknown functions in the form
 \be \label{1}
\left( {h^i}_\mu \right)=  \left( \matrix{ \mathbb{N}  &0 &0
 \vspace{3mm} \cr  0   & \mathbb{N}_1  &0  \vspace{3mm} \cr r\mathbb{N}_2   & 0  & r  \cr }
\right)\; , \ee
 where $\mathbb{N}(r)$, $\mathbb{N}(r)_1$ and $\mathbb{N}(r)_2$ are three  unknown functions.
The   metric spacetime of  triad (\ref{1}) takes the form
\begin{equation} ds^2=(\mathbb{N}^2-r^2\mathbb{N}_2{}^2)dt^2-\mathbb{N}_1{}^2dr^2-r^2d\phi^2-2r^2\mathbb{N}_2d{\phi}dr.\end{equation}
 Using Eq. (\ref{1}) in Eq. (\ref{Tor_sc}), we get
 \begin{equation} \label{2}
{{\textsl T}}=\frac{4\mathbb{N}\mathbb{N}'+r^3\mathbb{N}'_2{}^2}{2r\mathbb{N}^2\mathbb{N}_1{}^2}, \qquad \mbox{where} \qquad \mathbb{N}'=\frac{d\mathbb{N}}{dr}.\end{equation}
Now we are going to study the two separate cases of the field Eqs. (\ref{q8}).\vspace{0.2cm}\\

\subsection{ The vacuum (non-charged) case}
 Applying the triad (\ref{1}) to  Eq (\ref{q8}),  when ${\Theta}^\nu_\mu=0$,   we get
 \begin{eqnarray} \label{3}
& & I^t{}_t=\frac{(2\mathbb{N}^2+r^3\mathbb{N}_2\mathbb{N}'_2)f_{TT} T'}{r\mathbb{N}^2\mathbb{N}_1{}^2}+\frac{f_T}{r\mathbb{N}^3\mathbb{N}_1{}^3}\Biggl(r^3\mathbb{N}\mathbb{N}_1\mathbb{N}_2\mathbb{N}''_2+r^3\mathbb{N}\mathbb{N}_1\mathbb{N}'_2{}^2 -r^2\mathbb{N}_2\mathbb{N}'_2[r\mathbb{N}\mathbb{N}'_1+\mathbb{N}_1(r\mathbb{N}'-3\mathbb{N})]\nonumber\\
& & -2\mathbb{N}^3\mathbb{N}'_1+2\mathbb{N}^2\mathbb{N}_1\mathbb{N}'\Biggr)-f+2\Lambda=0,\nonumber\\
& & I^t{}_\phi=  \frac{(2r\mathbb{N}\mathbb{N}_2\mathbb{N}'-r^3\mathbb{N}_2{}^2\mathbb{N}'_2-r\mathbb{N}_2{}^2\mathbb{N}'_2-2\mathbb{N}^2\mathbb{N}_2)f_{TT} T'}{r\mathbb{N}^2\mathbb{N}_1{}^2}-\frac{f_T}{r\mathbb{N}^3\mathbb{N}_1{}^3}\Biggl(r\mathbb{N}\mathbb{N}_1(r^2\mathbb{N}_2{}^2+\mathbb{N}^2)\mathbb{N}''_2-2r\mathbb{N}^2\mathbb{N}_1\mathbb{N}_2\mathbb{N}'' \nonumber\\
& & +2r^3\mathbb{N}\mathbb{N}_1\mathbb{N}_2\mathbb{N}'_2{}^2-\mathbb{N}'_2(r^2\mathbb{N}_2{}^2+\mathbb{N}^2)[r\mathbb{N}\mathbb{N}'_1+\mathbb{N}_1(r\mathbb{N}'-3\mathbb{N})]+2\mathbb{N}^2\mathbb{N}_2\mathbb{N}'_1[r\mathbb{N}'-\mathbb{N}]\Biggr)=0,\nonumber\\
 \nonumber\\
& &I^r{}_r=\frac{f_T(4\mathbb{N}\mathbb{N}'+r^3\mathbb{N}'_2{}^2)}{r\mathbb{N}^2\mathbb{N}_1{}^2}-f+2\Lambda=0,  \nonumber\\
& &I^\phi{}_t=\frac{r^2\mathbb{N}'_2f_{TT} T'}{\mathbb{N}^2\mathbb{N}_1{}^2}+\frac{rf_T(r\mathbb{N}\mathbb{N}_1\mathbb{N}''_2-\mathbb{N}'_2[r\mathbb{N}\mathbb{N}'_1+\mathbb{N}_1(r\mathbb{N}'-3\mathbb{N})])}{\mathbb{N}^3\mathbb{N}_1{}^3}=0,  \nonumber\\
& &I^\phi{}_\phi=\frac{[2N\mathbb{N}'-r^2\mathbb{N}_2\mathbb{N}'_2]f_{TT} T'}{\mathbb{N}^2\mathbb{N}_1{}^2}-\frac{f_T}{r\mathbb{N}^3\mathbb{N}_1{}^3}\Biggl(r^3\mathbb{N}\mathbb{N}_1\mathbb{N}_2\mathbb{N}''_2-2r\mathbb{N}^2\mathbb{N}_1\mathbb{N}''+r^3\mathbb{N}\mathbb{N}_1\mathbb{N}'_2{}^2-r^2\mathbb{N}_2\mathbb{N}'_2\Biggl[r\mathbb{N}_1\mathbb{N}'\nonumber\\
& &+\mathbb{N}(r\mathbb{N}'_1-3\mathbb{N}_1)\Biggr]+2\mathbb{N}{}^2\mathbb{N}'[r\mathbb{N}'_1-\mathbb{N}_1]\Biggr)-f+2\Lambda=0, \nonumber\\
 \end{eqnarray}
 where $\mathbb{N}'=\frac{d \mathbb{N}(r)}{dr}$,  $\mathbb{N}'_1=\frac{d \mathbb{N}_1(r)}{dr}$, $\mathbb{N}'_2=\frac{d \mathbb{N}_2(r)}{dr}$.
Using the quadratic form of $f(T)$, i.e., $f(T)=T+\epsilon T^2$ in Eq. (\ref{3}) we get
 \begin{eqnarray} \label{4}
& & I^t{}_t= \frac{2\epsilon(2\mathbb{N}^2+r^3\mathbb{N}_2\mathbb{N}'_2)T'}{r\mathbb{N}^2\mathbb{N}_1{}^2}+\frac{(1+2\epsilon T)}{r\mathbb{N}^3\mathbb{N}_1{}^3}\Biggl(r^3\mathbb{N}\mathbb{N}_1\mathbb{N}_2\mathbb{N}''_2+r^3\mathbb{N}\mathbb{N}_1\mathbb{N}'_2{}^2 -r^2\mathbb{N}_2\mathbb{N}'_2[r\mathbb{N}\mathbb{N}'_1+\mathbb{N}_1(r\mathbb{N}'-3\mathbb{N})]\nonumber\\
& & -2\mathbb{N}^3\mathbb{N}'_1+2\mathbb{N}^2\mathbb{N}_1\mathbb{N}'\Biggr)-(T+\epsilon T^2)+2\Lambda=0,\nonumber\\
& & I^t{}_\phi= \frac{2\epsilon(2r\mathbb{N}\mathbb{N}_2\mathbb{N}'-r^3\mathbb{N}_2{}^2\mathbb{N}'_2-r\mathbb{N}_2{}^2\mathbb{N}'_2-2\mathbb{N}^2\mathbb{N}_2) T'}{r\mathbb{N}^2\mathbb{N}_1{}^2}-\frac{(1+2\epsilon T)}{r\mathbb{N}^3\mathbb{N}_1{}^3}\Biggl(r\mathbb{N}\mathbb{N}_1(r^2\mathbb{N}_2{}^2+\mathbb{N}^2)\mathbb{N}''_2-2r\mathbb{N}^2\mathbb{N}_1\mathbb{N}_2\mathbb{N}'' \nonumber\\
& & +2r^3\mathbb{N}\mathbb{N}_1\mathbb{N}_2\mathbb{N}'_2{}^2-\mathbb{N}'_2(r^2\mathbb{N}_2{}^2+\mathbb{N}^2)[r\mathbb{N}\mathbb{N}'_1+\mathbb{N}_1(r\mathbb{N}'-3\mathbb{N})]+2\mathbb{N}^2\mathbb{N}_2\mathbb{N}'_1[r\mathbb{N}'-\mathbb{N}]\Biggr)=0,\nonumber\\
& &I^r{}_r=\frac{(1+2\epsilon T)(4\mathbb{N}\mathbb{N}'+r^3\mathbb{N}'_2{}^2)}{r\mathbb{N}^2\mathbb{N}_1{}^2}-(T+\epsilon T^2)+2\Lambda=0,  \nonumber\\
& &I^\phi{}_t= \frac{2\epsilon r^2\mathbb{N}'_2T'}{\mathbb{N}^2\mathbb{N}_1{}^2}+\frac{r(1+2\epsilon T)(r\mathbb{N}\mathbb{N}_1\mathbb{N}''_2-\mathbb{N}'_2[r\mathbb{N}\mathbb{N}'_1+\mathbb{N}_1(r\mathbb{N}'-3\mathbb{N})])}{\mathbb{N}^3\mathbb{N}_1{}^3}=0,  \nonumber\\
& &I^\phi{}_\phi= \frac{2\epsilon[2N\mathbb{N}'-r^2\mathbb{N}_2\mathbb{N}'_2]T'}{\mathbb{N}^2\mathbb{N}_1{}^2}-\frac{(1+2\epsilon T)}{r\mathbb{N}^3\mathbb{N}_1{}^3}\Biggl(r^3\mathbb{N}\mathbb{N}_1\mathbb{N}_2\mathbb{N}''_2-2r\mathbb{N}^2\mathbb{N}_1\mathbb{N}''+r^3\mathbb{N}\mathbb{N}_1\mathbb{N}'_2{}^2-r^2\mathbb{N}_2\mathbb{N}'_2\Biggl[r\mathbb{N}_1\mathbb{N}'\nonumber\\
& &+\mathbb{N}(r\mathbb{N}'_1-3\mathbb{N}_1)\Biggr]+2\mathbb{N}{}^2\mathbb{N}'[r\mathbb{N}'_1-\mathbb{N}_1]\Biggr)-(T+\epsilon T^2)+2\Lambda=0, \nonumber\\
 \end{eqnarray}
 where $T'=\displaystyle\frac{dT}{dr}$.  It is  interesting  to note that if the dimensional parameter $\epsilon=0$ then Eq. (\ref{4}) reduces to that derived in \cite{Nijgmmp}.    Now we are going to solve the above system of differential equations using the following constrains $\Lambda=\frac{1}{24\epsilon}$ \cite{GSV}
\begin{eqnarray} \label{5}
\A \A i) \; \mathbb{N}=\frac{\sqrt{r^2-12c_1 \epsilon}}{\sqrt{12\epsilon}}, \qquad \qquad \mathbb{N}_1=\pm \frac{\sqrt{12 \epsilon}}{\sqrt{12c_1 \epsilon-r^2}}, \qquad \qquad \mathbb{N}_2=c_2, \nonu
\A \A ii)\; \mathbb{N}=\pm  \frac{\sqrt{r^4-12c_1\epsilon r^2+12c_3 {}^2\epsilon}}{r\sqrt{12\epsilon }}, \qquad  \qquad
 \mathbb{N}_1=\pm \frac{r\sqrt{12\epsilon }}{\sqrt{12c_1\epsilon r^2-r^4-12c_3{}^2 \epsilon}}, \qquad \qquad \mathbb{N}_2=c_2+\frac{c_3}{r^2}, \nonu
 \end{eqnarray}
 where $c_1$, $c_2$ and $c_3$ are constants of integration. It is clear that when the constant $c_3=0$ the second set of solution (\ref{5}) reduces to the first set of (\ref{5}).
 All the above sets of solution (\ref{5}) give   constant torsion, i.e., $T=\frac{1}{6\left|\epsilon\right|}$ which coincides with \cite{GSV}.  It is important to mention here  that  solution (\ref{5}) can not reduce to TEGR and therefore it has no analog in GR.
\subsection{The charged case}
 Applying the field Eq. (\ref{q8}) to triad (\ref{1}) we get the following  non-vanishing components when  $\Theta^\nu_\mu\neq0$
\begin{eqnarray} \label{6}
& &I^t{}_t=  \frac{(2\mathbb{N}^2+r^3\mathbb{N}_2\mathbb{N}'_2)f_{TT} T'}{r\mathbb{N}^2\mathbb{N}_1{}^2}+\frac{f_T}{r\mathbb{N}^3\mathbb{N}_1{}^3}\Biggl(r^3\mathbb{N}\mathbb{N}_1\mathbb{N}_2\mathbb{N}''_2+r^3\mathbb{N}\mathbb{N}_1\mathbb{N}'_2{}^2 -r^2\mathbb{N}_2\mathbb{N}'_2[r\mathbb{N}\mathbb{N}'_1+\mathbb{N}_1(r\mathbb{N}'-3\mathbb{N})]\nonumber\\
& & -2\mathbb{N}^3\mathbb{N}'_1+2\mathbb{N}^2\mathbb{N}_1\mathbb{N}'\Biggr)-f+2\Lambda- \frac{2q'^2}{\mathbb{N}^2\mathbb{N}_1{}^2}=0\nonumber\\
& & I^t{}_\phi=   \frac{(2r\mathbb{N}\mathbb{N}_2\mathbb{N}'-r^3\mathbb{N}_2{}^2\mathbb{N}'_2-r\mathbb{N}_2{}^2\mathbb{N}'_2-2\mathbb{N}^2\mathbb{N}_2)f_{TT} T'}{r\mathbb{N}^2\mathbb{N}_1{}^2}-\frac{f_T}{r\mathbb{N}^3\mathbb{N}_1{}^3}\Biggl(r\mathbb{N}\mathbb{N}_1(r^2\mathbb{N}_2{}^2+\mathbb{N}^2)\mathbb{N}''_2-2r\mathbb{N}^2\mathbb{N}_1\mathbb{N}_2\mathbb{N}'' \nonumber\\
& & +2r^3\mathbb{N}\mathbb{N}_1\mathbb{N}_2\mathbb{N}'_2{}^2-\mathbb{N}'_2(r^2\mathbb{N}_2{}^2+\mathbb{N}^2)
[r\mathbb{N}\mathbb{N}'_1+\mathbb{N}_1(r\mathbb{N}'-3\mathbb{N})]+2\mathbb{N}^2\mathbb{N}_2\mathbb{N}'_1
[r\mathbb{N}'-\mathbb{N}]\Biggr)+\frac{4\mathbb{N}_2{}^2q'^2}{\mathbb{N}^2\mathbb{N}_1{}^2}=0,\nonumber\\
 \nonumber\\
& &I^r{}_r= \frac{f_T(4\mathbb{N}\mathbb{N}'+r^3\mathbb{N}'_2{}^2)}{r\mathbb{N}^2\mathbb{N}_1{}^2}-f+2\Lambda- \frac{2q'^2}{\mathbb{N}^2\mathbb{N}_1{}^2}=0,  \nonumber\\
& & I^\phi{}_t= \frac{r^2\mathbb{N}'_2f_{TT} T'}{\mathbb{N}^2\mathbb{N}_1{}^2}+\frac{rf_T(r\mathbb{N}\mathbb{N}_1\mathbb{N}''_2-\mathbb{N}'_2[r\mathbb{N}\mathbb{N}'_1+\mathbb{N}_1(r\mathbb{N}'-3\mathbb{N})])}{\mathbb{N}^3\mathbb{N}_1{}^3}=0,  \nonumber\\
& &I^\phi{}_\phi= \frac{[2N\mathbb{N}'-r^2\mathbb{N}_2\mathbb{N}'_2]f_{TT} T'}{\mathbb{N}^2\mathbb{N}_1{}^2}-\frac{f_T}{r\mathbb{N}^3\mathbb{N}_1{}^3}\Biggl(r^3\mathbb{N}\mathbb{N}_1\mathbb{N}_2\mathbb{N}''_2-2r\mathbb{N}^2\mathbb{N}_1\mathbb{N}''+r^3\mathbb{N}\mathbb{N}_1\mathbb{N}'_2{}^2-r^2\mathbb{N}_2\mathbb{N}'_2\Biggl[r\mathbb{N}_1\mathbb{N}'\nonumber\\
& &+\mathbb{N}(r\mathbb{N}'_1-3\mathbb{N}_1)\Biggr]+2\mathbb{N}{}^2\mathbb{N}'[r\mathbb{N}'_1-\mathbb{N}_1]\Biggr)-f+2\Lambda+\frac{2q'^2}{\mathbb{N}^2\mathbb{N}_1{}^2}=0, \nonumber\\
 \end{eqnarray}
 where the unknown q(r) is  the electric charge which is defined  as \[A_\mu= q(r) \delta_\mu{}^t.\]
We mention here that the above charged differential equation of $f(T)$ are different from those of \cite{GSV} even when $\mathbb{N}_2=0$. The difference  raises due to the fact that the two field equations are different and become identical only when $f(T)=T$.  Equation (\ref{6}) reduces to (\ref{4}) when the unknown function $q(r)$ vanishing.  The general solutions of the above system of differential equations  using the same constrain of the uncharged case, i.e., ${\rm f(T)=T+\epsilon T^2}$ and $1-24\epsilon \Lambda=0$, take the following form
 \begin{eqnarray} \label{7}
\A \A i) \; \mathbb{N}=\frac{\sqrt{r^2-12c_1 \epsilon}}{\sqrt{12\epsilon}}, \qquad \qquad \mathbb{N}_1=\pm \frac{\sqrt{12 \epsilon}}{\sqrt{12c_1 \epsilon-r^2}}, \qquad \qquad \mathbb{N}_2=c_2,   \qquad q(r)=c_4,\nonu
\A \A ii)\; \mathbb{N}=\pm  \frac{\sqrt{r^4-12c_1\epsilon r^2+12c_3 {}^2\epsilon}}{r\sqrt{12\epsilon }}, \qquad
 \mathbb{N}_1=\pm \frac{r\sqrt{12\epsilon }}{\sqrt{12c_1\epsilon r^2-r^4-12c_3{}^2 \epsilon}}, \qquad  \mathbb{N}_2=c_2+\frac{c_3}{r^2}, \nonu
\A \A  \qquad q(r)=c_4,\nonu
\A \A iii)\; \mathbb{N}=\pm  \frac{c_{5} \mathbb{N}_3}{\sqrt{2r}},   \qquad
 \mathbb{N}_1=\pm \frac{2c_{5}(c_{5}r-1)\sqrt{3\epsilon}}{\mathbb{N}_3\sqrt{r}},  \qquad \mathbb{N}_2=c_{6}, \qquad  q(r)=c_{4}+c_{5}{}^2\ln(r)+\frac{c_{5}}{r},\nonu
 \A \A {\textrm with}  \qquad \qquad \mathbb{N}_3=\sqrt{4c_{5}{}+12r\epsilon c_{4}+2r[1+3\ln(r)]c_{5}{}^2-c_{5}{}^4r^3},
 \end{eqnarray}
where $c_4$, $c_5$ and $c_6$ are  integration constants.  It is necessary  to mention here that the above black hole solutions cannot reduce to that derived in \cite{GSV} due to the appearance of the constant $c_{5}$. This constant cannot be equal to zero otherwise we get a travail charge and return to the non-charged case given by (\ref{5}). This leads us to say that the charged solution derived in \cite{GSV} is not a black hole  solution of the present  $f(T)$ theory because the charged term must have the logarithmic  term as  in (\ref{7}) in addition to the monopole like one. A final remark about solution (\ref{7}) is that the logarithmic term,    which  appears in the potential,  is  not standard    in the Einstein-Maxwell (2+1)-dimensional theory \cite{BLM15}. Therefore, solution (\ref{7}) is a new analytic black hole solution in the frame  of $f(T)$ gravitational theory  whose field equations are  given by (\ref{q8}) and when $f(T)=T+\epsilon T^2$. In the next section, we are going to extract the physics of the uncharged and charged solutions by calculating their metrics, singularities and their energies.
 \newsection{Black hole physics}
Now we are going to discuss the physical meaning of the above black hole solutions considering the main features of the related black holes.\vspace{0.2cm}\\
\subsection{The non-charged metric}
The metric of the first set  of solution (\ref{5}) has the form
\be \label{8} ds_1{}^2=\frac{(r^2-12c_1 \left|\epsilon\right|-12c_2{}^2\left|\epsilon\right| r^2)}{12\left|\epsilon\right|}dt^2-\frac{12 \left|\epsilon\right| dr^2}{r^2-12c_1  \left|\epsilon\right|}-r^2d\phi^2-2r^2c_2dtd\phi.\ee
We can eliminate the cross term that appears in Eq. (\ref{8}) using the following transformation
\be \label{9} c_2t+\phi \rightarrow \phi' .\ee Using Eq. (\ref{9}) in  (\ref{8})  we get
\be \label{10} ds_1{}^2=\frac{(r^2-12c_1 \left|\epsilon\right|)}{12\left|\epsilon\right|}dt'^2-\frac{12 \left|\epsilon\right| dr^2}{r^2-12c_1 \left|\epsilon\right|}-r^2d\phi'^2.\ee
Equation (\ref{10}) can be rewritten as
\be\label{11} ds_1{}^2=(r^2\Lambda_e-c_1) dt'^2-\frac{ dr^2}{r^2\Lambda_e-c_1}-r^2d\phi'^2,\ee
where $\Lambda_e=\frac{1}{12\left|\epsilon\right|}.$ Equation (\ref{11}) shows that the metric asymptotes to AdS/dS.
For the second set  of solution (\ref{5}) the metric takes the form
\be \label{12}  ds_2{}^2=\frac{ r^2- 12\left|\epsilon\right| c_1-12c_2\left|\epsilon\right|[ c_2 r^2+2c_3]}{12 \left|\epsilon\right|}dt^2-\frac{12 r^2 \left|\epsilon\right|dr^2}{r^4-12c_1  \left|\epsilon\right|r^2+12c_3{}^2 \left|\epsilon\right|}-r^2d\phi^2-2(r^2c_2+c_3)dtd\phi.\ee
The cross term in Eq. (\ref{12}) can not be removed  by a coordinate transformation due to the appearance of the constant $c_3$. This constant is responsible for the rotating term which comes from the unknown function $\mathbb{N}_3$. Eq. (\ref{12}) can be rewritten as
\be \label{13} ds_2{}^2=(r^2\Lambda_e-  c_1)dt^2-\frac{ r^2 dr^2}{r^4\Lambda_e-c_1r^2+c_3{}^2}-r^2d\phi^2-2c_3dtd\phi,\ee where we have put the constants $c_2=0$. Again Eq. (\ref{13}) asymptotically goes to  AdS/dS solution.
\\
\\
\subsection{The metric of charged case}
The first two sets of the charged solution  (\ref{7}) are the same as  the two sets  of the non-charged solution. The metric of the third set of Eq. (\ref{7}) takes the form
\ba \A \A \label{14}
  ds_3{}^2=\frac{r^3[c_{5}{}^6-2c_{6}{}^2]-4c_{5}{}^3-12 r \epsilon c_{4}c_{5}{}^2-2rc_{5}{}^4-6r c_{5}{}^4 \ln r }{2r}dt^2\nonu
  \A \A   -\frac{12c_{5}{}^2 \epsilon (rc_{5}-1)^2}{r(r^3 c_{5}{}^4-4c_{5}-12 r \epsilon c_{4}-2rc_{5}{}^2-6r c_{5}{}^2 \ln r)}dr^2-r^2d\phi^2-2r^2c_{6}dtd\phi.\ea
  Using the following transformation
  \[c_6t+\phi \rightarrow \phi', \]  we can eliminate the cross term that appears in Eq. (\ref{14}) and get
 \ba \A \A \label{15}
  ds_3{}^2=\frac{r^3c_{5}{}^6-4c_{5}{}^3-12 r \left|\epsilon\right| c_{4}c_{5}{}^2-2rc_{5}{}^4-6r c_{5}{}^4 \ln r }{2r}dt^2-r^2d\phi'^2\nonu
  \A \A   -\frac{12c_{5}{}^2 \left|\epsilon\right| (r c_{5}-1)^2}{r(r^3 c_{5}{}^4-4c_{5}-12 r \left|\epsilon\right|c_{4}-2rc_{5}{}^2-6r c_{5}{}^2 \ln r)}dr^2.\ea
Eq. (\ref{15}) can be rewritten as
 \ba \label{16} \A \A
  ds_3{}^2=\left( r^2\Lambda_e-\frac{2}{r\sqrt{6\left|\epsilon\right|}}-\frac{1+(6\epsilon)^{4/3}c_4+3\ln r}{\sqrt[3]{36\epsilon^2}}\right)dt^2 -\frac{1}{f\left(r^2 \Lambda_e-\frac{2}{r\sqrt{6\left|\epsilon\right|}}-\frac{1+(6\epsilon)^{4/3}c_4+3\ln r}{\sqrt[3]{36\epsilon^2}}\right)}dr^2-r^2d\phi'^2,\nonu
  \A \A\ea
  where $c_5=\sqrt[6]{2\Lambda_e}$ and $f=\displaystyle\frac{1}{(1-\frac{1}{rc_5})^2}$. The metric of Eq. (\ref{16}) asymptotes AdS/dS spacetime. It interesting to note that, from Eq. (\ref{16}), we cannot recover  Eq. (\ref{11}). This is due to the fact that the third set of solutions (\ref{7}) cannot return to the first set.

  The torsion scalar of the non-charged case,  given by solution (\ref{5}), has the form
\be T_1=T_2=\frac{1}{6\left|\epsilon\right|},\ee
and, for the charged solution given by the third set of Eq. (\ref{7}), has the form
\be T_3=\frac{c_{5} r+2}{6r c_{5}\left|\epsilon\right|}.\ee


Now, let us discuss the singularities and the horizons of solution  (\ref{5}).  The curvature scalars arise from the metric of first set of solution (\ref{5}) have the form
 \begin{eqnarray} \label{17} \A \A R^{\mu \nu \lambda \rho}R_{\mu \nu \lambda \rho} =-R^{\mu \nu}R_{\mu \nu}=-\frac{1}{12\epsilon^2}, \qquad \qquad R =\frac{1}{2\left|\epsilon\right|}, \nonu
  \A\A T^{\mu \nu \lambda}T_{\mu \nu \lambda} = \frac{r^4-12r^2\left|\epsilon\right| c_1+72\epsilon^2c_1{}^2}{3r^2\left|\epsilon\right|(r^2-12\left|\epsilon\right| c_1)}\sim\left(\frac{1}{r^4}\right), \nonu
\A \A T^\mu T_\mu =  \frac{(r^2-6c_1\left|\epsilon\right|)^2}{3r^2\left|\epsilon\right|(r^2-12c_1\left|\epsilon\right|)}\sim\left(\frac{1}{r^4}\right).
\end{eqnarray} For the second set of solution (\ref{5}),   the invariants of curvature do not change from the first set however, the invariants of the torsion are given by
\begin{eqnarray}
T^{\mu \nu \lambda}T_{\mu \nu \lambda} \A=\A -\frac{72\epsilon^2c_1{}^2r^4-12\left|\epsilon\right| c_1r^6+r^8+144\epsilon^2c_1c_3{}^2 r^2-24\left|\epsilon\right| c_3{}^2r^4-144\epsilon^2c_3{}^4}
{3r^4\left|\epsilon\right|(12\left|\epsilon\right|[c_1r^2-c_3{}^2]-r^4)}\sim\left(\frac{1}{r^4}\right), \nonu
T^\mu T_\mu \A=\A  -\frac{12\left|\epsilon\right|(c_1r^2-2 c_3)^2}{r^4(12\left|\epsilon\right|[c_1r^2-c_3{}^2]-r^4)}\sim\left(\frac{1}{r^4}\right),\end{eqnarray}
 It is worth  mentioning here that the above invariants of the torsion behaves asymptotically as $\left(\frac{1}{r^4}\right)$ in contrast to the asymptotic  behavior in the TEGR case of the same triad which behaves as $\left(\frac{1}{r}\right)$. This  means that, in the $f(T)$  theory,   the invariants of  torsion go to zero fast than those of TEGR as $r \rightarrow \infty$. This means that the singularities of the invariants of $f(T)$ are much softer than those of TEGR. It is worth  to study if the solution of Eq. (\ref{5}) is stable or not by studying its anti-evaporation \cite{NO14}. All these issues need more investigation which will be clarified elsewhere.
The following scalars are satisfied for the two sets of solution (\ref{5}).
\be T(r)=\frac{1}{6\left|\epsilon\right|},\qquad \nabla_\alpha T^\alpha=\frac{1}{3\left|\epsilon\right|}, \qquad \Rightarrow R=-T-2\nabla_\alpha T^\alpha.
\ee
For the charged case, the invariants of the first two sets are not change however for the third set of solution (\ref{7}) we get the following invariants:
\begin{eqnarray}
\A \A  R^{\mu \nu \lambda \rho}R_{\mu \nu \lambda \rho} =\frac{1}{36r^2\epsilon^2c_{5}{}^4(c_{5}r-1)^6}\Biggl(3r^8c_{5}{}^{10}-10r^7c_{5}{}^9+3r^6c_{5}{}^8+12r^5c_{5}{}^7\ln r+24r^5\left|\epsilon\right|c_{5}{}^5c_{4}\nonu
\A \A +22r^5c_{5}{}^7-24r^4c_{5}{}^6\ln r-48\left|\epsilon\right| r^4c_{5}{}^4c_{4}+9r^4c_{5}{}^6-36r^3c_{5}{}^5\ln r+36r^2c_{5}{}^4(\ln r)^2-72r^3\left|\epsilon\right| c_{5}{}^3c_{4}\nonu
\A \A -108r^3c_{5}{}^5+144r^2\left|\epsilon\right|c_{5}{}^2 c_{4}  \ln r+84r^2c_{5}{}^4 \ln r+144r^2\epsilon^2 c_{4}{}^2+168r^2\left|\epsilon\right| c_{5}{}^2c_{4}+102r^2c_{5}{}^4\nonu
\A \A+24rc_{5}{}^3\ln r +48r \left|\epsilon\right|  c_{5} c_{4}-8rc_{5}{}^3+12c_{5}{}^2\Biggr)\sim\left(\frac{1}{r}\right), \nonu
\A \A  R^{\mu \nu}R_{\mu \nu}=\frac{1}{72r^2\epsilon^2c_{5}{}^4(c_{5}r-1)^6}\Biggl(6r^8c_{5}{}^{10}-20r^7c_{5}{}^9+5r^6c_{5}{}^8+24r^5c_{5}{}^7\ln r+48r^5\left|\epsilon\right| c_{5}{}^5c_{4}\nonu
\A \A +50r^5c_{5}{}^7-42r^4c_{5}{}^6\ln r-84\left|\epsilon\right| r^4c_{5}{}^4c_{4}-25r^4c_{5}{}^6-36r^3c_{5}{}^5\ln r+36r^2c_{5}{}^4(\ln r)^2-72r^3\left|\epsilon\right| c_{5}{}^3c_{4}\nonu
\A \A -88r^3c_{5}{}^5+144r^2\left|\epsilon\right| c_{5}{}^2 c_{4}  \ln r+90r^2c_{5}{}^4 \ln r+144r^2\epsilon^2 c_{4}{}^2+180r^2\left|\epsilon\right| c_{5}{}^2c_{4}+77r^2c_{5}{}^4\nonu
\A \A+24rc_{5}{}^3\ln r +48r \left|\epsilon\right|  c_{5} c_{4}+12rc_{5}{}^3+8c_{5}{}^2\Biggr)\sim\left(\frac{1}{r}\right), \nonu
\A \A R =\frac{3r^4c_{5}{}^5-5r^3c_{5}{}^4-3r^2c_{5}{}^3+6rc_{5}{}^2 \ln r+12r\left|\epsilon\right| c_{4}+8r c_{5}{}^2+2c_{5}}{6r\left|\epsilon\right| c_{5}{}^2(c_{5}r-1)^3}\sim\left(\frac{1}{r}\right),\nonu
\A \A T^{\mu \nu \lambda}T_{\mu \nu \lambda} =-\frac{-1}{6rc_{5}{}^2 \left|\epsilon\right| (rc_{5}-1)^2 (4c_{5}+2rc_{5}{}^2+12\left|\epsilon\right| r c_{4}+6r c_{5}{}^2 \ln r-r^3c_{5}{}^4)}\Biggl(2r^6c_{5}{}^8-12 r^4 c_{5}{}^6 \ln r\nonu
\A \A-24r^4 \left|\epsilon\right| c_{5}{}^4 c_{4}-10 r^4 c_{5}{}^6+36 r^2 c_{5}{}^4 (\ln r)^2 -4r^3 c_{5}{}^5+144 r^2 \left|\epsilon\right| c_{5}{}^2 c_{4} \ln r+24 r^2 c_{5}{}^4 \ln r+144r^2\epsilon^2 c_{4}{}^2\nonu
\A \A+48r^2\left|\epsilon\right| c_{5}{}^2 c_{4} +13 r^2 c_{5}{}^4+48 r c_{5}{}^3 \ln r +96 r \left|\epsilon\right| c_{5} c_{4} +4rc_{5}{}^3+20 c_{5}{}^2\Biggr)\sim \left(\frac{1}{r}\right), \nonu
\A \A T^\mu T_\mu = -\frac{(2c_{5}+5rc_{5}{}^2+12 r \left|\epsilon\right| c_{4}+6 r c_{5}{}^2 \ln r-2 r^3 c_{5}{}^4)^2}{12rc_{5}{}^2 \left|\epsilon\right| (rc_{5}-1)^2 (4c_{5}+2rc_{5}{}^2+12\left|\epsilon\right| r c_{4}+6r c_{5}{}^2 \ln r-r^3c_{5}{}^4)}\sim \left(\frac{1}{r}\right)\nonu
\A \A T(r)= \frac{c_{5} r+2}{6r c_{5} \left|\epsilon\right|}\sim \left(\frac{1}{r}\right), \qquad   \nabla_\alpha T^\alpha=\frac{13c_{5}{}^2+12 \left|\epsilon\right| c_{4}+6 c_{5}{}^2 \ln r -6r c_{5}{}^3-6r^2 c_{5}{}^4+4r^3 c_{5}{}^5}{6c_{5}{}^2 \left|\epsilon\right|(rc_{5}-1)^3 }\sim \left(\frac{1}{r}\right), \nonu
\A \A \Rightarrow R=-T-2\nabla_\alpha T^\alpha.\nonu
\end{eqnarray}
 The above invariants (non-charged and charged) show:\vspace{0.1cm}\\
 a)- All scalars of curvature and torsion show irregular behaviors when $r\rightarrow 0$ that describes a real singularity, except the invariants of curvature of the first and second sets of solution (\ref{5}). \vspace{0.1cm}\\
  b)- When the constant $c_1=\frac{r^2}{12\left|\epsilon\right|}$, we get a singular metric for the  first set of solution (\ref{5})  in addition  the invariants  $T^{\mu \nu \lambda}T_{\mu \nu \lambda}$ and $T^{\mu }T_{\mu}$ have a singular behavior and when $c_1=\frac{12c_3{}^2\left|\epsilon\right|+r^4}{12\left|\epsilon\right| r^2}$,  we get a singular metric for  the second  set of solution (\ref{5}). For the charged case, solution (\ref{7}), we get a singular metric when  $c_{4}=\frac{r^3 c_{5}{}^4-4c_{5}-2rc_{5}{}^2-6r c_{5}{}^2 \ln r}{12r \left|\epsilon\right|}$ .\vspace{0.1cm}\\
 c)-  For the third set of solution (\ref{7}), the scalars of curvature behave as  $\left(\frac{1}{r}\right)$ in contrast to TEGR or GR which behave as $\left(\frac{1}{r^2}\right)$.  However, the asymptotic behavior of the scalars of torsion does not change from the TEGR. \vspace{0.1cm}\\

 It is worth  mentioning that the above discussion shows that the dimensional parameter $\epsilon$ cannot be vanishing which ensures that all the above solutions have no analogue in GR.

\subsection{The Energy}
  In this subsection, we are going to carry out  the calculations of   black hole energy  solutions (\ref{5}) and (\ref{7})\footnote{We assume  gravitational coupling to have the form $G_{eff}=\frac{G_{Newton}}{1+f_T} $ \cite{WQM}. }.
From Eq. (\ref{q11}),  using the non-charged solution (\ref{5}),  we get
\be \label{18} S^{001}=-\frac{1}{2 r},  \ee using (\ref{18}) in (\ref{q11}),  we get
 \be
 P^0=E=-\frac{\pi (12c_1 \epsilon -r^2)}{27\kappa_3 \epsilon},\ee which is not finite. Therefore, to obtain a finite value of Eq. (\ref{q11}),  we use the following regularized expression
  \be \label{19} P^a:=\frac{1}{\kappa_3}\int_\Sigma d^2x  \partial_\nu\left[h{S}^{a 0
\nu} f(T)_T\right]-\frac{1}{\kappa_3}\int_\Sigma d^2x  \partial_\nu\left[h{S}^{a 0
\nu} f(T)_T\right].\ee Using (\ref{19}) in (\ref{5}), we get
\be \label{20} E=M,\ee
where $c_1=-\frac{9\kappa_3 M}{4\pi }$. The same above algorithm  can be applied to the second set of Eq. (\ref{5}) and the same value of Eq. (\ref{20}) can be obtained. For the third set of solution (\ref{7}) we get, after regularization, the energy in the form of (\ref{20}) up to $\left(\frac{1}{r}\right)$.

\section{Thermodynamics  }\label{S11}

 Hawking's  temperature of any solution can be  derived by requiring the
singularity  to disappear at the horizon using the Euclidean continuation method.
The  temperature of the outer event horizon at $r = r_h$, for the first set of solution (\ref{7}) as
\be \label{21}
T=\frac{1}{4\pi}\left(\frac{d g_{tt}(r)}{dr}\right)_{r_h}=\frac{ r_h}{24\pi\left|\epsilon\right|},\ee
and the  temperature of  the second set of solution (\ref{7})  is the same as that of Eq. (\ref{21}). Finally, the temperature of  the third set of solution (\ref{7}) has the form
\be \label{22}
T=\frac{ r_h{}^3-3r_h\sqrt[3]{36\epsilon^2}+2\sqrt[3]{6\epsilon}}{24\pi\left|\epsilon\right| r_h{}^2}.\ee

Now we are going to carry on the calculations of the entropy of the black hole solutions (\ref{5}) and (\ref{7}). For this purpose, we use the terminology
studied in \cite{MLM}.  The  entropy associated with
 any  solutions in the framework of $f(T)$ gravitational has the form \cite{MLM}
\be \label{23}
S=\frac{A}{4}f_T\mid_{r=r_h},\ee
where A is the horizon area. Using  the first and second sets of solution (\ref{7}) in (\ref{23}) we get
\be \label{24}  S= \pi r_{h}{}^2 \Biggl[1+2\left|\epsilon\right| T(r)\mid_{r=r_h}\Biggr]=\frac{4\pi r_{h}{}^2}{3} , \ee and for the third solution of Eq. (\ref{7}) we get
\be \label{25} S=\pi r_{h}{}^2\Biggl[1+\frac{r_{h}\sqrt[6]{6\left|\epsilon\right|}+2}{3r_{h}\sqrt[6]{6\left|\epsilon\right|}}\Biggr] . \ee
To investigate if the validity of the first law of the black hole solutions (\ref{7}) is satisfied or not we are going  to  discuss the paper by Miao et al. \cite{MLM}.  They \cite{MLM} rewrote the field equations (\ref{q8}), which are non-symmetric, into a symmetric part as well as a skew symmetric one as
\begin{eqnarray}\label{q88}
\A \A I_{(\mu\nu)}\stackrel {\rm def.}{=}S_{\mu \nu \rho} \partial^{\rho} T f_{TT}+f_T \left[G_{\mu \nu} -\frac{1}{2}g_{\mu \nu}T\right]
+\frac{f-2\Lambda}{2}g_{\nu \mu} =\kappa_3 {\cal T}_{\nu \mu},\nonu
\A \A I_{[\mu \nu]}\stackrel {\rm def.}{=} S_{[\mu \nu] \rho} \partial^{\rho} T f_{TT}=0.\end{eqnarray}
They have assumed  a Killing vector field whose  heat flux ${\rm \delta Q}$  has the form
\be \label{q89} {\rm \delta Q}=\frac{\kappa_3}{2\pi}\left[\frac{f_T dA}{4}\right]^{d\lambda}_0+\frac{1}{\kappa_3}\int_H k^\nu  f_{TT} \ T_{,\mu}(\xi^\rho S_{\rho \nu}{}^ \mu-\nabla_\nu \xi^\mu),\ee where $H$ refers to the black hole horizon.

In  \cite{MLM}, it is  proven that $\left[\frac{f_T dA}{4}\right]^{d\lambda}_0$ is equivalent to $T\delta S$ \cite{ MLM}. Therefore, if $\int_H k^\nu  f_{TT} \ T_{,\mu}(\xi^\rho S_{\rho \nu}{}^ \mu-\nabla_\nu \xi^\mu) \neq 0$,  then the first law of thermodynamics is violated.  It is shown that   the first law is always violated in $f(T)$ for non-trivial value of the  scalar torsion  \cite{MLM}.  In fact, the first and second sets of  solution (\ref{7}) have a trivial value of the scalar torsion and thus the first law is valid. However, the third set of black hole solution  (\ref{7}) has  a non-trivial value of the torsion scalar in addition that this solution is reproduced from the quadratic form of $f(T)$. Therefore, for this black hole solution,   the third set of solution (\ref{7}) violates the first law of thermodynamics.

\newsection{Concluding remarks}
In this paper, we have studied 3-dimensional $f(T)$   and  Maxwell-$f(T)$  gravity to check the existence of
circularly symmetric solutions.  To this end, we have applied  off diagonal triad having three unknown functions of the radial coordinate, to the field equations of  $f(T)$ theory (non-charged case).   We have solved the field equations exactly for the quadratic form  of $f(T)$ and have assumed  the following relation between the cosmological constant and the dimension parameter $\epsilon$, i.e., $\Lambda=\frac{1}{24 \epsilon}$ to simplify the form of the solution.  We have obtained analytic solution having two sets which can be categorized as:\vspace{0.1cm}\\
 i) The first set makes the off-diagonal component has a constant value.\vspace{0.1cm}\\
 ii) The second one has a non-trivial value of the off-diagonal component.\vspace{0.1cm}\\
 All of these sets are new  and have no analogues in standard GR because of the existence  of the dimensional parameter $\epsilon$, coefficient of the quadratic term of the scalar torsion. Such a parameter cannot  be equal to zero, otherwise, we get a singular form of the torsion scalar as well as of the metric.   All these sets give  constant torsion, i.e., $T=\frac{-1}{6\epsilon}$. The singularities of these sets have been studied and we have indicated that all the scalars derived  from curvature tensor  as well as from torsion tensor   show a singularity if the dimensional parameter $\epsilon$.   The asymptotic  behavior of the scalars, constructed out from the torsion behaves as $\left(\frac{1}{r^4}\right)$ in contrast to what is derived both  in GR and TEGR \cite{BTZ,Nijgmmp}.  Finally, we have calculated the energy of these sets and shown that it does not depend on  the dimensional parameter $\epsilon$.

For the charged case we have applied the same triad  to the equation of Maxwell-$f(T)$ gravitation theory.   We have solved the resulted differential equations and obtained a solution which is a new one and completely  different from that derived in \cite{GSV}.  This  solution cannot reduce to that derived in  \cite{GSV} because of the difference of the field equations of the two theories.  As it is clear from the potential vector-like term, i.e. $ q(r)=c_{4}+c_{5}{}^2\ln(r)+\frac{c_{5}}{r}$,  if the constant  $c_5=0$,  we return to the first set of  solution (\ref{q4}) which is different the result presented in \cite{GSV}. It is interesting  to mention here that the
3-dimensional vector potential-like term  derived in GR (TEGR) depends only on the logarithm,  however our solution of higher order gravity (ultraviolet)  depends on an additional   monopole term.  We may consider this term as a correction due to the higher order gravity.

For the charged solution we have shown that the torsion scalar is not constant. Therefore,  this solution in higher-order torsion gravity  is completely different from GR (TEGR).

Finally, we have calculated some of the thermodynamical quantities like the Hawking temperature and the entropy.  For the non-charged sets, we have shown that the first law of thermodynamics is valid. However,   the charged case shows that the first law is not satisfied. The violation of the first law of thermodynamics comes form  the fact that the torsion scalar is not trivial and also $f_{TT}=2\epsilon$, i.e., it is not TEGR where $f(T)=T$ \cite{MLM}. This case needs more accurate  studies because of the non-trivial value of the scalar torsion  which is responsible for the deviation from TEGR due to the  non-vanishing value of $f_{TT}=\epsilon$.
In a forthcoming paper, a systematic discussion of thermodynamical properties of these solutions, in view of a (3+1) generalization, will be pursued.

\begin{acknowledgments}
GN is supported in part by the Egyptian Ministry of Scientific Research under project No. 24-2-12  and SC is  supported in part by the INFN sezione di Napoli, {\it iniziative specifiche} TEONGRAV and QGSKY. The  article is  based upon work from COST action CA15117 (CANTATA),
supported by COST (European Cooperation in Science and Technology).
\end{acknowledgments}


\begin{thebibliography}{99}

\bibitem{Ra}  A. G. Riess et al.  [Supernova Search Team Col- laboration], {\it Astron. J.} {\bf116} (1998), 1009  [astro- ph/9805201].

 \bibitem{Ps} S. Perlmutter et al.  [Supernova Cosmology Project Collaboration], {\it Astrophys. J.} {\bf 517} (1999), 565  [astro-ph/9812133].

\bibitem{Hg}  G. Hinshaw et al.  [WMAP Collabora- tion], {\it Astrophys. J. Suppl.} {\bf 208} (2013), 19  [arXiv:1212.5226 [astro-ph.CO]].

 \bibitem{Ed}  D. J. Eisenstein et al.  [SDSS Collaboration],  {\it Astrophys. J.}  {\bf 633} (2005), 560  [astro-ph/0501171].

\bibitem{Wy}  Y. Wang, {\it Phys. Rev.}  {\bf D 78} (2008), 123532 [arXiv:0809.0657 [astro-ph]].

\bibitem{Pp}   P. J. E. Peebles and B. Ratra, {\it Rev. Mod. Phys.}  {\bf 75}  (2003), 559 [astro-ph/0207347].

\bibitem{BT}  J. Binney, S. Tremaine, {\rm  Galactic Dynamics,  Princeton University Press, Princeton (1987)}


\bibitem{Be}   E. Berti et al., {\it Class. Quantum Grav.}  {\bf 32} (2015), 243001.

\bibitem{CF}   S. Capozziello and M. Francaviglia, {\it Gen. Rel. Grav.} {\bf 40} (2008), 357  [arXiv:0706.1146 [astro-ph]].

\bibitem{CDL} S. Capozziello and M. De Laurentis, {\it Phys.Rept.} {\bf  509} (2011) 167
[arXiv:1108.6266 [gr-qc]].

\bibitem{NO} S. Nojiri and S.D. Odintsov, {\it Phys.Rept.} {\bf 505 }(2011) 59
 [arXiv:1011.0544 [gr-qc]].

\bibitem{NOO} S. Nojiri, S.D. Odintsov, V.K. Oikonomou, {\it Phys.Rept.} {\bf 692} (2017) 1
[ arXiv:1705.11098 [gr-qc]].


\bibitem{Wc}   R. Aldrovandi and J. G. Pereira, {\it Teleparallel Gravity: An Introduction,} Springer, Dordrecth, (2012),  http://www.ift.unesp.br/users/jpereira/tele.pdf.

\bibitem{N08}  G. G. L. Nashed, {\it Euro. Phys. J.} {\bf C 54} (2008), 2,  291 [ arXiv:0804.3285 [gr-qc]].

\bibitem{APB}  R. Aldrovandi, J. G. Pereira and K. H. Vu, Braz. {\it J. Phys.} {\bf  34} (2004), 1374 [gr-qc/0312008].

\bibitem{Mj}  J. W. Maluf, {\it Annalen Phys.} {\bf 525} (2013), 339 [arXiv:1303.3897 [gr-qc]].

\bibitem{Ea}   A. Einstein, {\it S.B. Preuss. Akad. Wiss.} (1925) 414-419  see the translation in [arXiv:physics/0503046[physics.hist-ph]].

\bibitem{N06}  G. G. L. Nashed, {\it Mod. Phys. Lett.} {\bf A 21} (2006), 29,  2241 [  arXiv:gr-qc/0401041 [gr-qc]].

\bibitem{N008}  G. G. L. Nashed, {\it Int. J. Mod. Phys.} {\bf A 21} (2006), 15,  3181 [  arXiv:gr-qc/0401041 [gr-qc]].

\bibitem{Ea28}    A. Einstein, {\it S.B. Preuss. Akad. Wiss.} (1928) 217- 221, see the translation in [arXiv:physics/0503046[physics.hist-ph]].

 \bibitem{Ea29} A. Einstein, {\it Math. Ann.} {\bf 102} (1929), 685, see the translation in [arXiv:physics/0503046[physics.hist-ph]].

 \bibitem{BF9} G. R. Bengochea and R. Ferraro, {\it Phys. Rev.} {\bf D 79} (2009), 124019.

\bibitem{noi} Y. Cai, S. Capozziello, M. De Laurentis, E. N. Saridakis, {\it  Rept. Prog. Phys.} {\bf  79} (2016)  106901, [arXiv:1511.07586 [gr-qc]].

\bibitem{FF}  R. Ferraro and F. Fiorini, {\it Phys. Rev.} {\bf D 78} (2008), 124019 [arXiv:0812.1981 [gr-qc]].

\bibitem{FF09}  F. Fiorini and R. Ferraro, {\it Int. J. Mod. Phys.} {\bf A 24} (2009), 1686  [arXiv:0904.1767 [gr-qc]].

\bibitem{CRC}  V.F. Cardone, N. Radicella, S. Camera, {\it Phys. Rev.} {\bf D 85} (2012), 124007  [arXiv:1204.5294 [astro-ph]].

\bibitem{Mr}  R. Myrzakulov, {\it Eur. Phys. J.} {\bf C 71} (2011), 1752 [arXiv:1006.1120 [gr-qc]].

\bibitem{Gw}  G. G. L. Nashed, {\it Chin. Phys.} {\bf B 19} (2010), 2,  020401 [arXiv:0910.5124  [gr-qc]].

\bibitem{Yr}  R. J. Yang, {\it Eur. Phys. J.} {\bf C 71} (2011), 1797 [arXiv:1007.3571 [gr-qc]].

\bibitem{Bg}  G.R. Bengochea, {\it Phys. Lett.}  {\bf B 695} (2011), 405- 411.

\bibitem{BGLL}  K. Bamba, C.Q. Geng, C.C. Lee, L.W. Luo, {\it JCAP} {\bf 1101} (2011), 021.

\bibitem{KA}  K. Karami and A. Abdolmaleki, {\it Res. Astron. Astrophys.} {\bf 13} (2013), 757 [arXiv:1009.2459 [gr- qc]].

\bibitem{DDS}  J. B. Dent, S. Dutta, E. N. Saridakis, {\it JCAP} {\bf 1101} (2011), 009.

 \bibitem{CCDS} Y. F. Cai, S.H. Chen, J.B. Dent, S. Dutta, E.N. Saridakis, {\it Class. Quant. Grav.}  {\bf 28} (2011), 215011.

 \bibitem{CCFR} S. Capozziello, V.F. Cardone, H. Farajollahi, A. Ravanpak, {\it Phys. Rev.}  {\bf D 84}  (2011),  043527.

\bibitem{KOS}  K. Bamba, S. D. Odintsov and D. Saez- Gomez, {\it Phys. Rev.} {\bf D 88} (2013),  084042 [arXiv:1308.5789 [gr-qc]].

\bibitem{CCR}  S. Camera, V. F. Cardone and N. Radicella, {\it Phys. Rev.} {\bf D 89} (2014), 083520.

\bibitem{Ngrg15}  G. L. Nashed, {\it Gen. Rel. Grav.} {\bf 47}  (2015), 75.

\bibitem{ZA}  M. Zubair and G. Abbas, {\it arXiv:1507.00247} [physics.gen-ph].

\bibitem{Wt}  T. Wang, {\it Phys. Rev.}  {\bf D 84} (2011),  024042 [arXiv:1102.4410 [gr-qc]].

\bibitem{FF11}  R. Ferraro and F. Fiorini, {\it Phys. Rev.}  {\bf D 84} (2011),  083518 [arXiv:1109.4209 [gr-qc]].

\bibitem{GSV}  P. A. Gonzalez, E. N. Saridakis and Y. Vasquez, {\it JHEP} {\bf 1207} (2012), 053 [arXiv:1110.4024 [gr- qc]].

\bibitem{CGSV13}  S. Capozziello, P. A. Gonzalez, E. N. Saridakis and Y. Vasquez, {\it JHEP} {\bf 1302} (2013),  039 [arXiv:1210.1098 [hep-th]].

\bibitem{ACN} A.M. Awad, S. Capozziello, G.G.L. Nashed, {\it JHEP} {\bf 1707} (2017) 136
[arXiv:1706.01773 [gr-qc]].

\bibitem{CDG} A. de la Cruz-Dombriz, P. K. S. Dunsby and D. Saez-Gomez, {\it JCAP} {\bf 1412} (2014), 048.

\bibitem{RHTMM}  M. E. Rodrigues, M. J. S. Houndjo, J. Tossa, D. Momeni and R. Myrzakulov, {\it JCAP} {\bf 1311} (2013), 024 [arXiv:1306.2280 [gr-qc]].

\bibitem{Ngrg13}  G. G. L. Nashed, {\it Gen. Rel. Grav.}  {\bf 45} (2013), 1887 [arXiv:1502.05219 [gr-qc]].

\bibitem{N10}  G. G. L. Nashed, {\it Astrophys. Space Sci.} {\bf 330} (2010), 173 [arXiv:1503.01379 [gr-qc]].

\bibitem{JRH}  E. L. B. Junior, M. E. Rodrigues and M. J. S. Houndjo, {\it JCAP} {\bf 10} (2015), 060.

\bibitem{FLR} G. Farrugia, J. L. Said and  M. L. Ruggiero, {\it Phys. Rev.} {\bf D 93} (2016), 104034.

\bibitem{BFG}  C. Bejarano, R. Ferraro and M. J. Guzm\'an,
{\it Eur. Phys. J.} {\bf C 75} (2015), 77  [arXiv:1412.0641 [gr-qc]].

\bibitem{Km}  M. {Kr\v{s}\v{s}\'{a}k}, {\it Eur.Phys.J.} {\bf C77} (2017), 44 [arXiv:1510.06676 [gr-qc]].

\bibitem{KS}   M. {Kr\v{s}\v{s}\'{a}k} and E. N. Saridakis, {\it Class. Quantum Grav.}  {\bf 33} (2016), 115009.

\bibitem{BMT}  C.G. Bohmer, A. Messa, N. Tamanini, {\it Class. Quantum Grav.}  {\it 28} (2011), 245020

\bibitem{DWM}  H. Dong, Y.-b Wang, X.-h Meng, {\it Eur. Phys. J.} {\bf C. 72} (2012), 2002 .

\bibitem{Nprd}  G.G.L. Nashed, {\it Phys.  Rev.}  {\bf D 88} (2013), 104034.

\bibitem{CADT}  A. Paliathanasis, S. Basilakos, E.N. Saridakis, S. Capozziello, K. Atazadeh, F. Darabi, M. Tsamparlis, {\it Phys. Rev.}  {\bf D 89} (2014), 104042.

\bibitem{LSB} B. Li, T. P. Sotiriou, and J. D. Barrow, {\it Phys. Rev.} {\bf D 83} (2011), 104017.

\bibitem{LSB1} B. Li, T. P. Sotiriou, and J. D. Barrow, {\it Phys. Rev.} {\bf D 83} (2016), 064035.

\bibitem{BC17} S. Bahamonde and  S. Capozziello , {\it Euro. Phys. J.} {\bf C 77} (2017), 107.

\bibitem{BTZ} M. Banados, C. Teitelboim and J. Zanelli, {\it Phys. Rev. Lett.} {\bf 69} (1992), 1849.

 \bibitem{Cs} S. Carlip, {\it Class. Quant. Grav.} {\bf 12} (1995), 2853.

\bibitem{Cs1} S. Carlip, {\it Class. Quant. Grav.} {\bf 22} (2005), R85.


\bibitem{RM}  S. F. Ross and R. B. Mann, {\it Phys. Rev.} {\bf D 47} (1993), 3319.

\bibitem{HW} G. T. Horowitz and D. L. Welch, {\it Phys. Rev. Lett.} {\bf 71} (1993), 328.

\bibitem{BHTZ} M. Banados, M. Henneaux, C. Teitelboim and J. Zanelli, {\it Phys. Rev.} {\bf D 48} (1993), 1506; {\it Phys.
Rev.} {\bf D 88} (2013), 069902.

\bibitem{MTZ} C. Martinez, C. Teitelboim and J. Zanelli, {\it Phys. Rev.} {\bf D 61} (2000), 104013.

\bibitem{Cg} G. Clement, {\it Phys. Lett.} {\bf B 367} (1996), 70.

\bibitem{HPPS} S. H. Hendi, B. Eslam Panah, S. Panahiyan and A. Sheykh, {\it Phys. Lett.}{\bf B767} (2017), 214.

\bibitem{Wr} R. Weitzenbock, {\it Invariance Theorie, Nordhoff, Gronin-
gen, 1923}.

\bibitem{US3} S. C. Ulhoa and E. P. Spaniol,  {\it IJMP} {\bf D22} (2013), 1350069.

\bibitem{MDTC} J. W. Maluf, J. F. da Rocha-neto, T. M. L. Toribio and K. H.
Castello-Branco,  {\it Phys.\ Rev.\ }{\bf D65} (2002), 124001.

\bibitem{Nijgmmp} G.G.L. Nashed, to appear in  in {\it Int. J. Geometric Modern Method Phys.} {\bf 14} (2017), 1750105.

\bibitem{BLM15}  G. Barnich, P.-H. Lambert and P. Mao, {\it Class. Quantum Grav. } {\bf 32} (2015), 245001.

\bibitem{NO14}  S. Nojiri and S. D. Odintsov, {\it Phys. Lett.. } {\bf B735} (2014), 376.

\bibitem{MLM} R.-X. Miao, M. Li, Y.-G. Miao, {\it JCAP} {\bf 11} (2011), 033.

\bibitem{WQM} H. Wei, H.-Yu Qi, X.-Peng Ma, {\it Eur. Phys. J.} {\bf C72} (2012) 2117.

\end{thebibliography}
\end{document}